\documentclass[aps,pra,twocolumn,amsmath,amssymb,showpacs]{revtex4}

\usepackage{graphicx}
\usepackage{dcolumn}
\usepackage{bm}

\begin{document}

\title{Resonant Magnetic Vortices}

\author{Yves D\'ecanini}
\email{decanini@univ-corse.fr}
\author{Antoine Folacci}
\email{folacci@univ-corse.fr}
\affiliation{
SPE, UMR CNRS 6134, Equipe Physique Semi-Classique (et) de la Mati\`ere Condens\'ee \\
Universit\'e de Corse, Facult\'e des Sciences, BP 52, 20250 Corte,
France}

\date{\today}

\begin{abstract}

By using the complex angular momentum method, we provide a
semiclassical analysis of electron scattering by a magnetic vortex
of Aharonov-Bohm-type. Regge poles of the $S$-matrix are
associated with surface waves orbiting around the vortex and
supported by a magnetic field discontinuity. Rapid variations of
sharp characteristic shapes can be observed on scattering cross
sections. They correspond to quasibound states which are
Breit-Wigner-type resonances associated with surface waves and
which can be considered as quantum analogues of acoustic
whispering-gallery modes. Such a resonant magnetic vortex could
provide a new kind of artificial atom while the semiclassical
approach developed here could be profitably extended in various
areas of the physics of vortices.

\end{abstract}


\maketitle

\section{Introduction}

Modern electronic technology is based on the control of the
current and, with this aim in view, artificial structures such as
quantum wires, dots and other nanoscaled devices have been
developed (see, for example, Ref.~\onlinecite{Ando98} and
references therein). In this context, the Aharonov-Bohm (AB)
effect \cite{AB59, Olariu85, Peshkin89} is nowadays often
involved: a magnetic flux threading a confined region with leads
is used to modify the electron transport behavior. With this in
mind, we study, in this paper, an artificial structure which
presents richer scattering properties than the ordinary flux lines
usually considered. Such a system could provide a new kind of
artificial atom.

Here, we consider the scattering of a spinless electron (mass $M$,
charge $-e$ and total energy $E$) by the magnetic vortex defined,
in the cylindrical coordinate system $(\rho ,\varphi ,z)$, by the
electromagnetic potential
\begin{equation}\label{1}
V=\left\{ \begin{array}{cl} +\infty & \mathrm{for} \ \rho < R_1, \\
0 & \mathrm{for} \ \rho > R_1,
\end{array}
\right.
\mathbf{A}=\left\{ \begin{array}{cl} \frac{\Phi}{2\pi\rho} \
\mathbf{e_\varphi} & \mathrm{for} \ \rho < R_2, \\
0 & \mathrm{for} \ \rho > R_2,
\end{array}
\right.
\end{equation}
with $R_1 < R_2$. In Eq.~(\ref{1}), the scalar potential describes
a hard core which prevents the electron from entering the region
$\rho < R_1$ while the vector potential describes, in the radial
gauge, the magnetic field $\mathbf{B}=(\Phi/2\pi \rho)
\left(\delta(\rho)- \delta(\rho-R_2)\right)\mathbf{e_z}.$ This
field consists of a flux line at $\rho = 0$ (the usual AB flux
line) and an infinitely thin magnetic field shell localized at
$\rho = R_2$. Inside the region $\rho < R_2$, the total flux is
given by $\Phi $ while outside this region it vanishes. Such a
vortex was first suggested by Aharonov in order to avoid the
ambiguities of the incoming wave function of the electron in the
AB scattering, ambiguities arising from the long range behavior of
the vector potential. It was studied by Liang in
Ref.~\onlinecite{Liang85}.

In this article, the scattering of the electron is analyzed from a
semiclassical point of view by using the complex angular momentum
(CAM) method. It should be noted that this method has been
extensively used in several domains of scattering theory since the
pioneering work of Watson \cite{Watson18} dealing with the
propagation and diffraction of radio waves around the earth (see
the monographs of Newton \cite{New82} and of Nussenzveig
\cite{Nus92} and references therein for various applications in
quantum mechanics, nuclear physics, electromagnetism, optics,
acoustics and seismology). As far as we know, the CAM method has
never been introduced in the context of the AB effect (see,
however, the end of Ref.~\onlinecite{Berry80} where Berry
envisages the possibility of such an approach) or in order to
study the electron scattering by a magnetic field as well as its
resonant aspects. As we shall show below, this semiclassical
approach permits us to emphasize the aspects of scattering linked
to time-reversal invariance breaking as well as the rich resonant
properties of the magnetic vortex and more particularly the
existence of quasibound states associated with surface waves
orbiting around the vortex and supported by the magnetic field
shell.

\section{Exact $S$-matrix and resonances}

From now on, we treat our problem in a two-dimensional setting,
ignoring the z coordinate. We denote by $\hat H$ the Hamiltonian
of the electron and we introduce its wave number
$k=\sqrt{2ME}/\hbar$ as well as the quantum flux parameter $\alpha
= -e\Phi / 2\pi\hbar$. We are first interested by the construction
of the $S$-matrix. Because of the cylindrical symmetry of the
vortex, the $S$-matrix is diagonal and its elements $S_{\ell m}$
are given by $S_{\ell m}=S_\ell \ \delta _{\ell m}$. For a given
angular momentum index $\ell \in \mathbf{Z}$, the coefficient
$S_\ell$ is obtained from the partial wave $\Psi _\ell$ solution
of the following problem \cite{Mott65}:
\begin{description}
  \item (i) $\Psi _\ell$ satisfies the time-independent Schr\"odinger
equation $(\hat H - E)\Psi _\ell=0$,
  \item (ii) $\Psi _\ell$ vanishes at $\rho = R_1$ (impenetrability
of the hard core of radius $R_1$) while $\Psi _\ell$ as well as
its normal derivative are both continuous at $\rho = R_2$ (which
guarantee the continuity of the probability density and of the
radial component of the probability current at $\rho = R_2$
\cite{Liang85}),
  \item (iii) at large distance, $\Psi _\ell$ has the asymptotic
behavior
\begin{eqnarray}
&&\Psi_\ell(\rho,\varphi) \underset{\rho \to +\infty}{\sim}
\frac{1}{\sqrt{2\pi k\rho}}\left(e^{-i(k\rho -\ell\pi/2-\pi/4)} \right. \nonumber \\
&& \qquad\qquad\qquad +   \left. S_\ell e^{i(k\rho
-\ell\pi/2-\pi/4)}\right)e^{i\ell\varphi}. \nonumber
\end{eqnarray}
\end{description}
In the region $R_1 < \rho < R_2$, $\hat H$ is the standard AB
Hamiltonian given by
\begin{equation}\label{HamAB}
{\hat H} =-\frac{\hbar^2}{2M} \left[ \frac{\partial ^2}{\partial
\rho^2}+\frac{1}{\rho}\frac{\partial }{\partial
\rho}+\frac{1}{\rho^2}\left( \frac{\partial }{\partial \varphi}
-i\alpha \right)^2 \right]
\end{equation}
and the solution of (i) is expressible in terms of Bessel
functions (see \cite{AS65}) as a linear combination of
$J_{\ell-\alpha}(k\rho)e^{i\ell\varphi}$ and
$H^{(1)}_{\ell-\alpha}(k\rho)e^{i\ell\varphi}$. Let us recall that
in the usual AB scattering, the Bessel function indices are
$|\ell-\alpha |$ with $\ell \in \mathbf{Z}$. Here, absolute values
are unnecessary because we do not require the regularity of the
modes at $\rho = 0$. In the region $\rho
> R_2$, $\hat H$ reduces to the free Hamiltonian
and the solution of (i) can be constructed from
$J_\ell(k\rho)e^{i\ell\varphi}$ and
$H^{(1)}_\ell(k\rho)e^{i\ell\varphi}$. As a consequence, the
partial wave $\Psi _\ell$  solution of (i) and (ii) can be
obtained exactly. Then, by using the standard asymptotic behavior
of Hankel functions \cite{AS65}, we find from (iii) the expression
of the diagonal elements $S_\ell$ of the $S$-matrix:
\begin{equation}
S_\ell(k)=1-2\frac{D^{(1)}_\ell(k)}{D_\ell(k)}
\end{equation}
where $D^{(1)}_\ell(k)$ and $D_\ell(k)$ are two $3\times 3$
determinants which are explicitly given by
\begin{subequations}\label{DD}
\begin{eqnarray}
&D^{(1)}_\ell(k)=n_{\ell-\alpha}(k)J_\ell(kR_2)
-d_{\ell-\alpha}(k)J'_\ell(kR_2), \qquad \ \\
&D_\ell(k)=n_{\ell-\alpha}(k)H_\ell^{(1)}(kR_2)-d_{\ell-\alpha}(k)H_\ell^{(1)'}(kR_2),\quad
\end{eqnarray}
with
\begin{eqnarray}
&&n_\mu(k)=H_\mu^{(1)}(kR_1)J'_\mu(kR_2)-J_\mu(kR_1)H_\mu^{(1)'}(kR_2), \\
&&d_\mu(k)=H_\mu^{(1)}(kR_1)J_\mu(kR_2)-J_\mu(kR_1)H_\mu^{(1)}(kR_2).
\qquad
\end{eqnarray}
\end{subequations}
The unitarity of the $S$-matrix \cite{New82}, which expresses the
probability conservation, can be easily verified by using
elementary properties of Bessel functions. The reciprocity
property \cite{New82}, which is associated with time-reversal
invariance, is not satisfied because $S_\ell$ is not an even
function of $\ell$. Furthermore, the $S$-matrix is not invariant
under the change $\alpha \to \alpha +1$. Here, and by contrast to
the case of the ordinary AB scattering, the electron can
distinguish between two fluxes which differ by an integer multiple
of $2\pi\hbar/e$.

From the $S$-matrix, one can construct the scattering amplitude
\begin{equation}\label{ampli}
f(k, \varphi)=\sqrt{\frac{1}{2i\pi k}}
\sum_{\ell=-\infty}^{+\infty} \left(S_\ell(k) - 1
\right)e^{i\ell\varphi}
\end{equation}
while the total scattering cross section can be obtained by using
the optical theorem:
\begin{equation}\label{CS}
\sigma _T(k)=\frac{1}{2\pi}\sqrt{\frac{8\pi}{k}} \mathrm{Im}
\left( e^{-i\pi/4}f(k, \varphi =0)\right).
\end{equation}

\begin{figure}
\includegraphics[height=6cm,width=8.6cm]{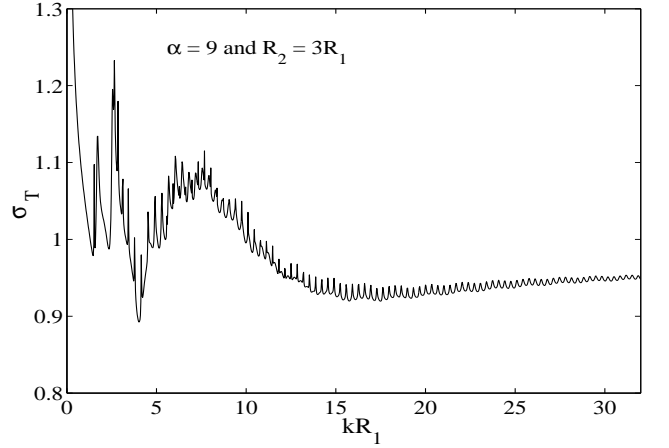}
\caption{\label{fig:cross} The normalized total cross section
$\sigma _T$. }
\end{figure}
\begin{figure}
\includegraphics[height=6cm,width=8.6cm]{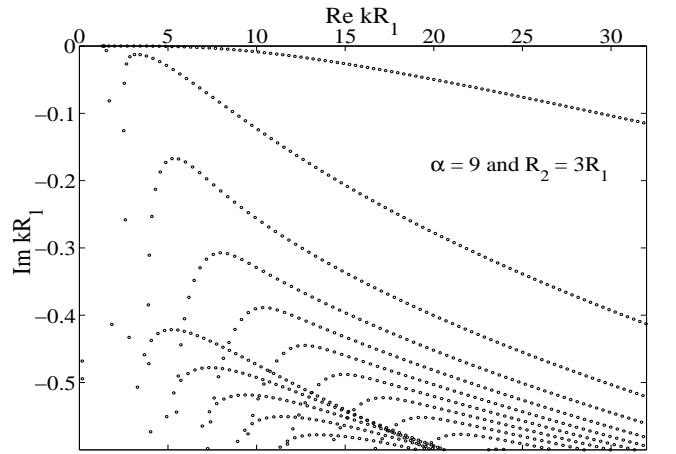}
\caption{\label{fig:exres} Scattering resonances in the complex
$kR_1$-plane.}
\end{figure}

\noindent In Fig.~\ref{fig:cross}, we present the total cross
section as a function of $kR_1$. As far as the numerical aspects
of our work are concerned, we restrict ourselves to the particular
configuration $\alpha = 9$ and $R_2 = 3R_1$ (we note, however,
that the results emphasized numerically are very general) and we
furthermore use the reduced wave number $kR_1$ instead of the wave
number $k$. Rapid variations of sharp characteristic shapes can be
observed. This strongly fluctuating behavior is due to scattering
resonances. These resonances are the poles of the $S$-matrix lying
in the fourth quadrant of the complex $k$-plane and they are
determined by solving
\begin{equation}\label{det}
D_\ell(k)=0 \quad \mathrm{for} \quad \ell \in \mathbf{Z}.
\end{equation}
The solutions of (\ref{det}) are denoted by $k_{\ell
p}=k^{(o)}_{\ell p}-i\Gamma _{\ell p}/2$ where $k^{(o)}_{\ell
p}>0$ and $\Gamma _{\ell p}>0$, the index $p$ permitting us to
distinguish between the different roots of (\ref{det}) for a given
$\ell$. In the immediate neighborhood of the resonance $k_{\ell
p}$, $S_\ell(k)$ has the Breit-Wigner form, i.e., is proportional
to
\begin{equation}\label{BW}
\frac{\Gamma _{\ell p}/2}{k-k^{(o)}_{\ell p}+i\Gamma _{\ell p}/2}.
\end{equation}
As a consequence, when a pole of the $S$-matrix is sufficiently
close to the real axis in the complex $k$-plane, it has an
appreciable influence on the total cross section. In
Fig.~\ref{fig:exres}, resonances are exhibited. A one-to-one
correspondence between the peaks of $\sigma _T$ in
Fig.~\ref{fig:cross} and the resonances near the real $kR_1$-axis
can be clearly observed. In the scattering of an electron with
wave number $k^{(o)}_{\ell p}$, a decaying state of the
electron-vortex system is formed. It has a finite lifetime
proportional to $1/\Gamma _{\ell p}$ and which increases with
$\alpha$. When the corresponding scattering resonance can be
observed on the total cross section, it is a long-lived state.
Because of these quasibound states, the magnetic vortex behaves as
a kind of artificial atom.

\section{Semiclassical analysis}

\begin{figure}
\includegraphics[height=6cm,width=8.6cm]{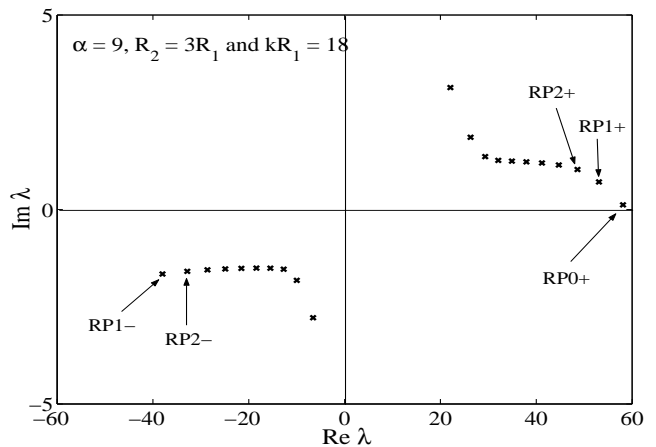}
\caption{\label{fig:RP} Regge poles in the complex angular
momentum plane.}
\end{figure}

Using the CAM method, we can provide a physical picture of the
scattering process in term of diffraction by surface waves and a
physical explanation of the mechanism of resonance excitation
valid for high wave numbers. We shall work from the scattering
amplitude but we are fully aware of the fact that a more rigorous
(but also more longer) analysis could be done from the trace of
the Green function. We first apply the Poisson summation formula
to the scattering amplitude (\ref{ampli}). We have
\begin{equation}\label{ampliII}
f(k, \varphi)=\sqrt{\frac{1}{2i\pi k}} \sum_{m=-\infty}^{+\infty}
\int_{-\infty}^{+\infty} d\lambda \left(S_\lambda(k) - 1
\right)e^{i\lambda(\varphi+m2\pi)}.
\end{equation}
We now go over the complex $\lambda $-plane. $\lambda $ is then
called the CAM and $S_\lambda(k)$ is then an analytical extension
of $S_\ell (k)$ into the CAM plane. We can then deform the path of
integration in (\ref{ampliII}) taking into account the possible
singularities. The only singularities that are met are the poles
of the S-matrix lying in the CAM plane. They are known as Regge
poles \cite{New82,Nus92} and are determined by solving
\begin{equation}\label{RP}
D_\lambda(k)=0 \quad \mathrm{for} \quad k > 0.
\end{equation}
The solutions of (\ref{RP}) are denoted by $\lambda_p$, the index
$p$ permitting us to distinguish between the different roots. We
can then extract from (\ref{ampliII}) the contribution of a
residue series over Regge poles given by
\begin{equation}\label{ampliIII}
f_S(k, \varphi)=\sqrt{\frac{2i\pi }{k}} \sum_{p} \pm r_p(k)
\sum_{m=1}^{+\infty}e^{i\lambda_p(k)(\varphi \pm m2\pi)}
\end{equation}
where $r_p(k)=\mathrm{residue}\left(S_\lambda(k)\right)_{\lambda =
\lambda _p(k)}$. Here the $+$ and $-$ signs are associated with
the Regge poles lying respectively in the first and in the third
quadrant of the CAM plane. It should be noted that $f$ differs
from $f_S$ by contour integrals which provide an uniform
approximation valid in a large range around $\varphi=0$. We are
not interested by such a contribution which does not play any role
in the resonance phenomenon. In Eq.~(\ref{ampliIII}), terms like
$\exp(i\lambda_p(k)(\varphi \pm m2\pi))$ are surface wave
contributions. A Regge pole lying in the first (resp. the third)
quadrant of the CAM plane corresponds to a surface wave
propagating counterclockwise (resp. clockwise) around the magnetic
vortex and $\mathrm{Re} \ \lambda_p(k)$ represents its azimuthal
propagation constant while $\mathrm{Im} \ \lambda_p(k)$ is its
damping constant. In (\ref{ampliIII}), we take into account the
multiple circumnavigations around the magnetic vortex.
Fig.~\ref{fig:RP} exhibits the distribution of Regge poles for a
given wave number. The Regge poles in the first and third
quadrants are not symmetrically distributed as a consequence of
the breaking of time-reversal invariance. The Regge pole
$\mathrm{RP}0+$ is very close to the real axis in the complex
$\lambda$-plane. It then corresponds to a surface wave which is
slightly attenuate during its propagation and which contributes
significantly to the scattering process and to the resonance
mechanism. The Regge pole $\mathrm{RP}1+$ is not so close to the
real axis but it could have an appreciable influence on the
resonance mechanism while the other Regge poles are too far to
contribute so significantly.

\begin{figure}
\includegraphics[height=6cm,width=8.6cm]{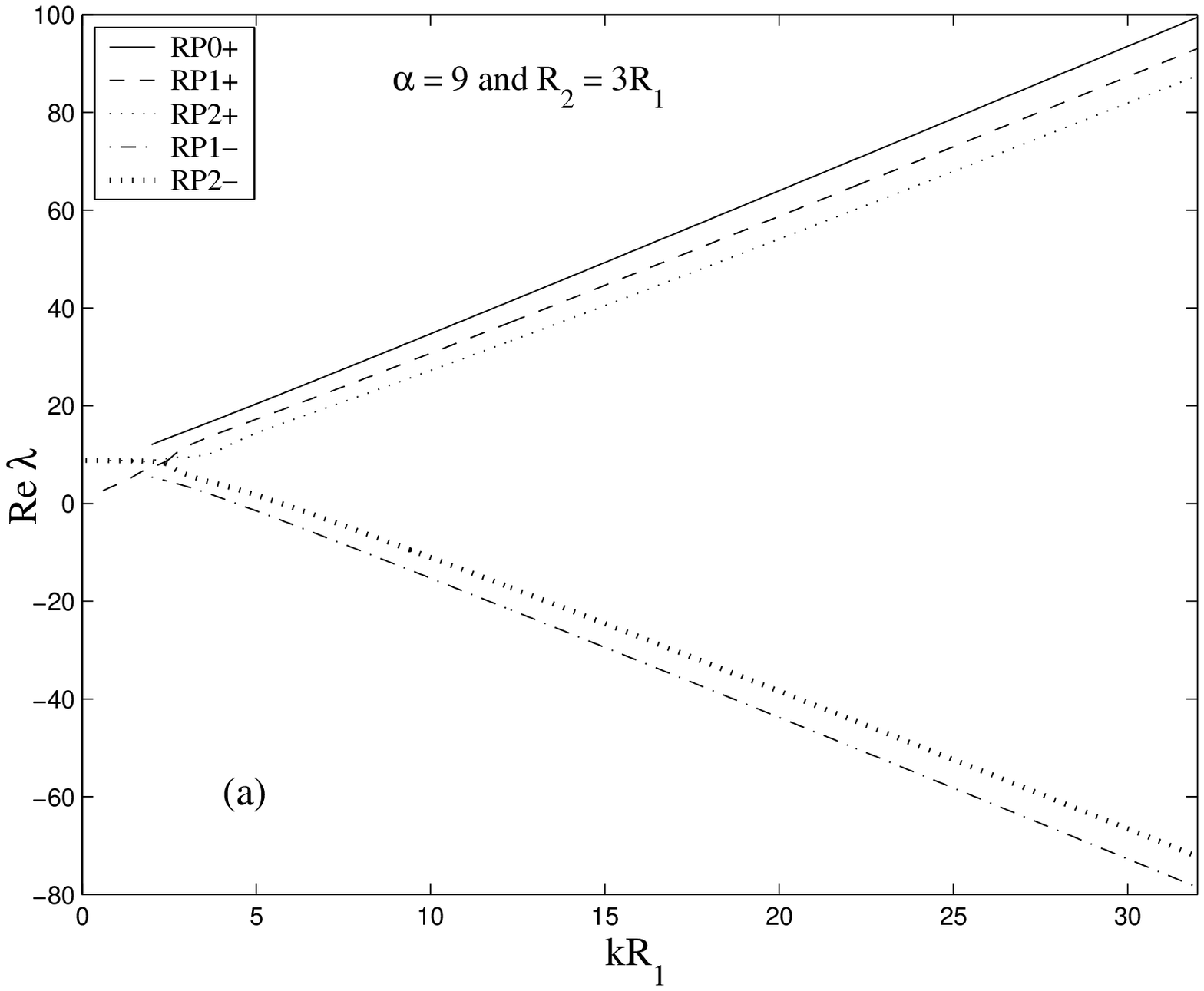}
\includegraphics[height=6cm,width=8.6cm]{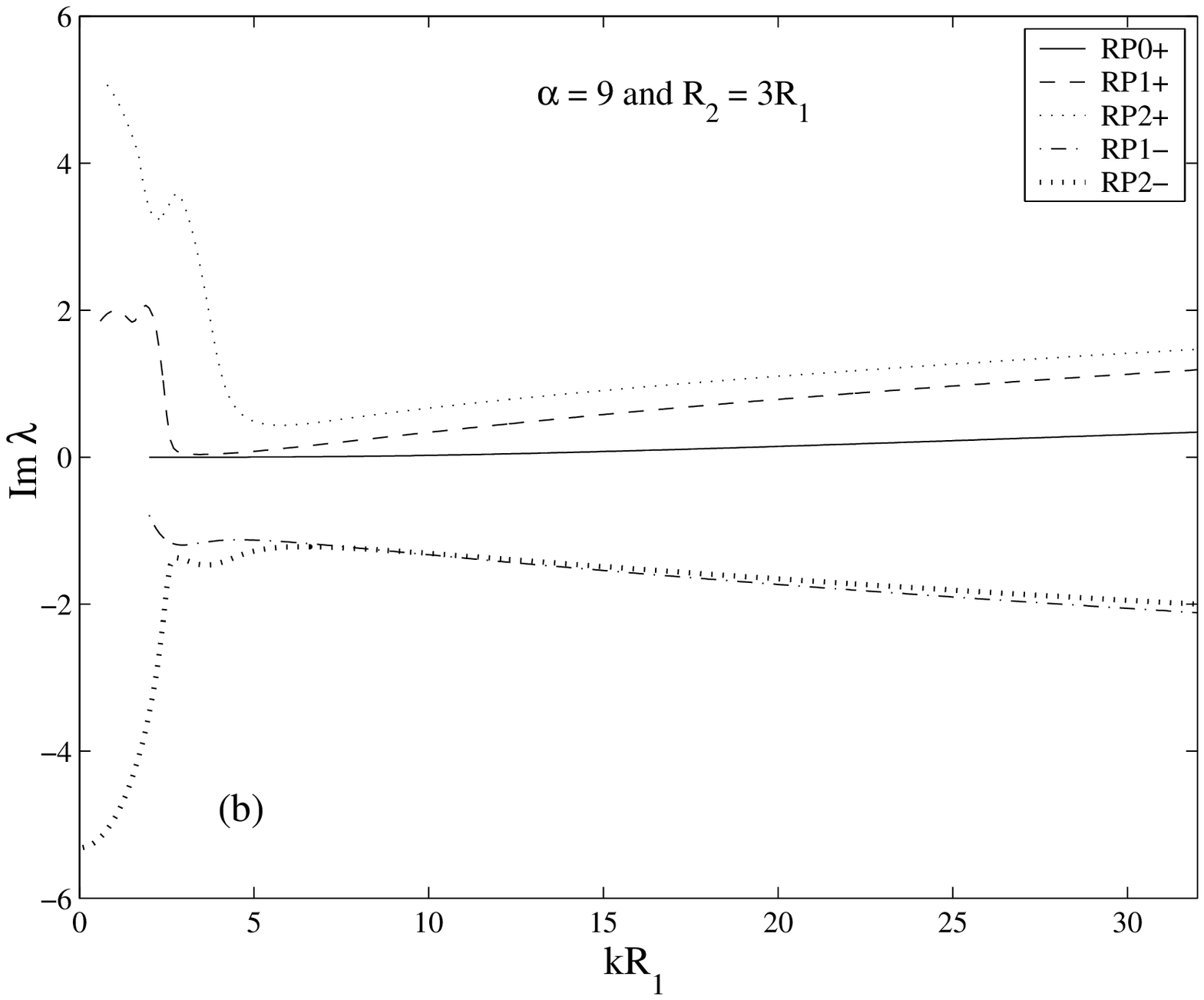}
\caption{\label{fig:RT}Regge trajectories.}
\end{figure}

\begin{figure}
\includegraphics[height=6cm,width=8.6cm]{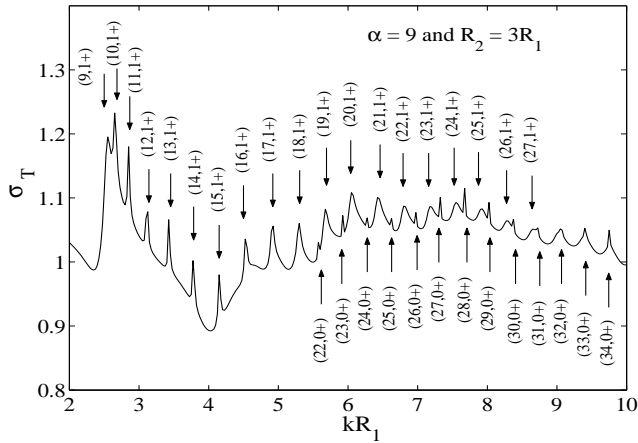}
\caption{\label{fig:resoncross} Identification of resonances on
the total cross section.}
\end{figure}

Resonance phenomenons (and the resonant behavior of $f_S$ and
therefore of $\sigma _T$) can now be understood in terms of
surface waves. As $k$ varies, each Regge pole $\lambda_p(k)$
describes a trajectory (usually designed as a Regge trajectory
\cite{New82}) in the CAM plane. When the quantity $\mathrm{Re} \
\lambda_p(k)$ coincides with an integer, a resonance occurs.
Indeed, it is produced by a constructive interference between the
different components of the pth surface wave, each component
corresponding to a different number of circumnavigations.
Resonance wave numbers are therefore obtained from the
Bohr-Sommerfeld quantization condition
\begin{equation}\label{sc1}
\mathrm{Re} \ \lambda_p \left(k^{(o)}_{\ell p}  \right)= \ell
\quad \mathrm{for} \quad \ell \in \mathbf{Z}.
\end{equation}
By assuming that $k$ is in the neighborhood of $k^{(o)}_{\ell p}$
and using $\mathrm{Re} \ \lambda_p (k ) \gg \mathrm{Im} \
\lambda_p (k)$ (which can be numerically verified, except for very
low frequencies),  we can expand $\lambda_p(k)$ in a Taylor series
about $k^{(o)}_{\ell p}$, and obtain
\begin{eqnarray} \label{TS}
\lambda_p(k)& \approx &\ell + \left(\frac{d \ \mathrm{Re} \
\lambda_p(k)}{dk}\right)_{k =k^{(o)}_{\ell p}} (k - k^{(o)}_{\ell
p}  ) \nonumber  \\ &  & \qquad + i \ \mathrm{Im} \ \lambda_p
(k^{(o)}_{\ell p}).
\end{eqnarray}
Then, after summation over $m$ in (\ref{ampliIII}), we find, by
using (\ref{TS}), that $f_S(k, \varphi)$ presents a resonant
behavior given by the Breit-Wigner formula (\ref{BW}) with
\begin{equation}\label{sc2}
\frac{\Gamma _{\ell p}}{2}=\left(  \frac{\mathrm{Im} \ \lambda_p
(k)}{d \ \mathrm{Re} \ \lambda_p (k) /dk}
\right)_{k=k^{(o)}_{\ell p}}.
\end{equation}
\begin{figure}
\includegraphics[height=6cm,width=8.6cm]{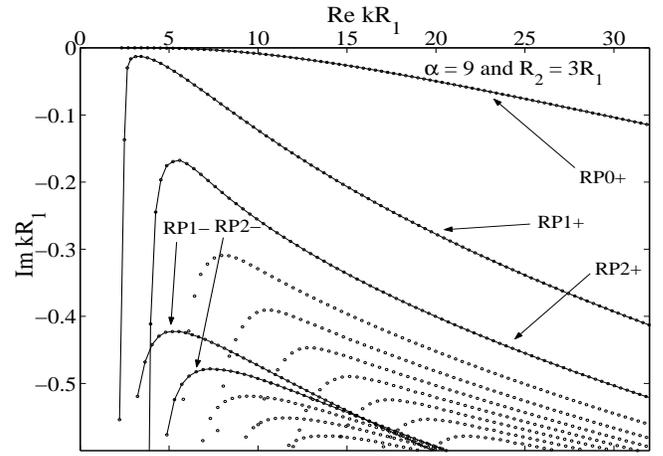}
\caption{\label{fig:scres} Semiclassical resonances in the complex
$kR_1$-plane.}
\end{figure}
\noindent Eqs.~(\ref{sc1}) and (\ref{sc2}) are semiclassical
formulas which permit us to determine the location of the
resonances from Regge trajectories. Fig.~\ref{fig:scres} exhibits
the resonance distribution obtained from the semiclassical
formulas (\ref{sc1}) and (\ref{sc2}) by using the Regge
trajectories numerically calculated (see Fig.~\ref{fig:RT}). A
comparison of Fig.~\ref{fig:scres} and Fig.~\ref{fig:exres} (or a
numerical comparison between the exact and the semiclassical
spectra) shows a very good agreement, except for low values of the
wave number $k$. Furthermore, as indicated in
Fig.~\ref{fig:scres}, the semiclassical theory permits us to
classify the resonances in distinct families, each family being
associated with one Regge pole and therefore to understand the
meaning of the index $p$ introduced to denote the solutions of
(\ref{det}). In Fig.~\ref{fig:resoncross}, a zoom in on
Fig.~\ref{fig:cross} is realized and the peaks are associated with
the semiclassical resonances linked to the Regge poles
$\mathrm{RP}0+$ and $\mathrm{RP}1+$.

A deeper understanding of the scattering process as well as its
dependence on the quantum flux parameter $\alpha$ can be obtained
by solving perturbatively Eq.~(\ref{RP}). Wait \cite{Wait}
presents a method in order to study acoustic whispering-gallery
phenomena in circular cylinders. Such a method is equally valid,
mutatis mutandis, for finding the solutions of (\ref{RP}). By
using in (\ref{DD}) the appropriate asymptotic expansions
\cite{AS65} (Debye expansions and uniform expansions in terms of
the Airy function) for Bessel functions, we can obtain an
asymptotic approximation for the Regge poles $\lambda _p(k)$:
\begin{subequations}\label{RPW}
\begin{eqnarray}
& & \lambda _{\mathrm{RP}0+}(k)= kR_2 + \alpha +
u_0^+(kR_2)\left(\frac{kR_2}{2}\right)^{1/3} + \dots  \qquad    \\
& & \lambda _{\mathrm{RP}p\pm}(k)= \pm kR_2 + \alpha  \pm
u_{p}^\pm(kR_2) \left(\frac{kR_2}{2}\right)^{1/3} + \dots \qquad
\end{eqnarray}
\end{subequations}
with $p=1,2, \dots $. Here $u_0^+(kR_2)$ and $u_p^\pm(kR_2)$ are
given by
\begin{eqnarray}
&&  u_0^+= t_0 + i  \frac{\ \ {\tau_0}^{1/2}}{(-t_0)}
\exp\left(-\frac{4}{3}{\tau_0}^{3/2}  \right) ,   \nonumber   \\
&&   u_p^\pm=t_p+i  \frac{{\ \ (-\tau_{p\pm})}^{1/2}}{(-t_p)},
\nonumber
\end{eqnarray}
where
\begin{equation}
t_0=-\left(\frac{3\pi}{4}\right)^{2/3}, \
t_p=-\left[\frac{3\pi}{2}\left( p+ \frac{1}{4} \right)
\right]^{2/3},    \nonumber
\end{equation}
and
\begin{equation}
\tau_0= t_0 + \alpha  \left(\frac{2}{kR_2} \right)^{1/3},    \
\tau_{p\pm}= t_p \pm \alpha  \left(\frac{2}{kR_2} \right)^{1/3}.
\nonumber
\end{equation}
In Table~\ref{tab:table1}, we present some results for Regge
poles. A comparison between the ``exact" and the asymptotic values
shows a rather good agreement. It could be possible to greatly
improve the imaginary parts of the asymptotic results by taking
into account higher orders of the perturbative series (\ref{RPW}).

\begin{table}
\caption{\label{tab:table1} A comparison between the exact and the
asymptotic values for Regge poles ($\alpha=9$, $R_2=3R_1$ and
$kR_1=18$).}
\begin{ruledtabular}
\begin{tabular}{ccc}
Regge poles &  $\lambda_p$ exact &   $\lambda_p$ semiclassical   \\
\hline
RP0+& +58.11+0.117i & +57.69+0.305i   \\
RP1+& +53.08+0.708i & +53.21+0.470i  \\
RP2+& +48.62+1.027i & +48.52+0.840i   \\
RP3+& +44.71+1.143i & +44.50+0.866i     \\
RP1--& -\ 38.02\ -\ 1.658i & -\ 35.21\ -\ 2.302i   \\
RP2--& -\ 32.83\ -\ 1.590i & -\ 30.52\ -\ 1.739i     \\
RP3--& -\ 28.62\ -\ 1.552i & -\ 26.50\ -\ 1.473i
\end{tabular}
\end{ruledtabular}
\end{table}

By replacing (\ref{RPW}) in (\ref{ampliIII}), it is obvious that
all the surface waves are supported by the infinitely thin
magnetic field shell localized at $\rho = R_2$.  Furthermore,
because of the term $\alpha $ which appears in (\ref{RPW}),
surface wave propagating around the magnetic vortex acquire a
geometrical phase $\alpha \varphi \pm \alpha m 2\pi$ (the + and -
signs correspond respectively to surface waves propagating
counterclockwise and clockwise) reminiscent of the Aharonov-Bohm
effect. Resonances are therefore associated with
whispering-gallery modes mainly concentrated at the inside surface
of the magnetic field discontinuity at $\rho = R_2$. Here we have
encountered, in the context of quantum mechanics, the
whispering-gallery phenomena described long time ago by Rayleigh
\cite{Rayleigh1887, Rayleigh1910} in acoustics.

\section{Conclusion and perspectives}

To conclude, we would like first to comment on some aspects of our
work and then to consider some possible directions which are
suggested by our results:

- We have investigated the resonant properties of a rather special
magnetic vortex for which the Schr\"odinger equation can be solved
exactly in terms of Bessel functions. As a consequence, we have
been able to perform the exact as well as the asymptotic
calculations involved. We think that scattering by such a vortex
cannot be considered as a thought experiment. Such a vortex could
be realized and experimentally studied because, in the domain of
the AB effect, experimentalists have developed ingenious
techniques (see, for example, Ref.~\onlinecite{Peshkin89}). At any
rate, even if that is not possible, we think that resonant
properties as well as surface waves could be experimentally
observed from more general magnetic vortices.

- The CAM method could be naturally used in many other areas of
the physics of vortices and in particular in order to study, from
a semiclassical point of view, scattering by vortices in
superfluids and Bose-Einstein condensates or in superconductors.
However, it seems to us that our approach could be above all
profitably extended in all the domains where analogs of the AB
effect have been recently developed, extending the pioneering
contribution of Berry \textit{et al} \cite{Berryetal80} (see
Refs.~\onlinecite{Roux97, Manneville01, Berthet01} for the
acoustical AB effect, Refs.~\onlinecite{Vivanco99, Coste99I,
Coste99II, Bernal02} for the hydrodynamical AB effect and
Ref.~\onlinecite{Neshev01} for the optical AB effect). In the
particular case of the scattering of ultrasonic waves by
hydrodynamic vortices, the CAM method could provide new
interpretations of sound and flow interactions in connection with
the geometrical theory of diffraction.

- Finally, applications in the domain of electronics and
mesoscopic physics could be envisaged and the resonant magnetic
vortex used to modify the electron transport behavior. Surface
waves as well as the associated resonances could not only
contribute significantly to conductance oscillations, but also
lead to new effects.

\begin{acknowledgments}
We thank Remy Berthet for fruitful discussions.
\end{acknowledgments}

\bibliography{RMVPRA}%

\begin{thebibliography}{23}
\expandafter\ifx\csname natexlab\endcsname\relax\def\natexlab#1{#1}\fi
\expandafter\ifx\csname bibnamefont\endcsname\relax
  \def\bibnamefont#1{#1}\fi
\expandafter\ifx\csname bibfnamefont\endcsname\relax
  \def\bibfnamefont#1{#1}\fi
\expandafter\ifx\csname citenamefont\endcsname\relax
  \def\citenamefont#1{#1}\fi
\expandafter\ifx\csname url\endcsname\relax
  \def\url#1{\texttt{#1}}\fi
\expandafter\ifx\csname urlprefix\endcsname\relax\def\urlprefix{URL }\fi
\providecommand{\bibinfo}[2]{#2}
\providecommand{\eprint}[2][]{\url{#2}}

\bibitem[{\citenamefont{Ando et~al.}(1998)\citenamefont{Ando, Arakawa, Furuya,
  Komiyama, and Nakashima}}]{Ando98}
\bibinfo{author}{\bibfnamefont{T.}~\bibnamefont{Ando}},
  \bibinfo{author}{\bibfnamefont{Y.}~\bibnamefont{Arakawa}},
  \bibinfo{author}{\bibfnamefont{K.}~\bibnamefont{Furuya}},
  \bibinfo{author}{\bibfnamefont{S.}~\bibnamefont{Komiyama}}, \bibnamefont{and}
  \bibinfo{author}{\bibfnamefont{H.}~\bibnamefont{Nakashima}},
  \emph{\bibinfo{title}{Mesoscopic Physics and Electronics}}
  (\bibinfo{publisher}{Springer-Verlag, Berlin}, \bibinfo{year}{1998}).

\bibitem[{\citenamefont{Aharonov and Bohm}(1959)}]{AB59}
\bibinfo{author}{\bibfnamefont{Y.}~\bibnamefont{Aharonov}} \bibnamefont{and}
  \bibinfo{author}{\bibfnamefont{D.}~\bibnamefont{Bohm}},
  \bibinfo{journal}{Phys.\ Rev.} \textbf{\bibinfo{volume}{115}},
  \bibinfo{pages}{485} (\bibinfo{year}{1959}).

\bibitem[{\citenamefont{Olariu and Popescu}(1985)}]{Olariu85}
\bibinfo{author}{\bibfnamefont{S.}~\bibnamefont{Olariu}} \bibnamefont{and}
  \bibinfo{author}{\bibfnamefont{I.~I.} \bibnamefont{Popescu}},
  \bibinfo{journal}{Rev.\ Mod.\ Phys.} \textbf{\bibinfo{volume}{57}},
  \bibinfo{pages}{339} (\bibinfo{year}{1985}).

\bibitem[{\citenamefont{Peshkin and Tonomura}(1989)}]{Peshkin89}
\bibinfo{author}{\bibfnamefont{M.}~\bibnamefont{Peshkin}} \bibnamefont{and}
  \bibinfo{author}{\bibfnamefont{A.}~\bibnamefont{Tonomura}},
  \emph{\bibinfo{title}{The Aharonov-Bohm Effect, {\rm Lecture Notes in
  Physics, Vol. 340}}} (\bibinfo{publisher}{Springer-Verlag, Berlin},
  \bibinfo{year}{1989}).

\bibitem[{\citenamefont{Liang}(1985)}]{Liang85}
\bibinfo{author}{\bibfnamefont{J.~Q.} \bibnamefont{Liang}},
  \bibinfo{journal}{Phys.\ Rev.\ D} \textbf{\bibinfo{volume}{32}},
  \bibinfo{pages}{1014} (\bibinfo{year}{1985}).

\bibitem[{\citenamefont{Watson}(1918)}]{Watson18}
\bibinfo{author}{\bibfnamefont{G.~N.} \bibnamefont{Watson}},
  \bibinfo{journal}{Proc.\ Roy.\ Soc.\ London A}
  \textbf{\bibinfo{volume}{100}}, \bibinfo{pages}{83} (\bibinfo{year}{1918}).

\bibitem[{\citenamefont{Newton}(1982)}]{New82}
\bibinfo{author}{\bibfnamefont{R.~G.} \bibnamefont{Newton}},
  \emph{\bibinfo{title}{Scattering Theory of Waves and Particles}}
  (\bibinfo{publisher}{Springer-Verlag, New-York}, \bibinfo{year}{1982}),
  \bibinfo{edition}{2nd} ed.

\bibitem[{\citenamefont{Nussenzveig}(1992)}]{Nus92}
\bibinfo{author}{\bibfnamefont{H.~M.} \bibnamefont{Nussenzveig}},
  \emph{\bibinfo{title}{Diffraction Effects in Semiclassical Scattering}}
  (\bibinfo{publisher}{Cambridge University Press, Cambridge},
  \bibinfo{year}{1992}).

\bibitem[{\citenamefont{Berry}(1980)}]{Berry80}
\bibinfo{author}{\bibfnamefont{M.}~\bibnamefont{Berry}},
  \bibinfo{journal}{Eur.\ J.\ Phys.} \textbf{\bibinfo{volume}{1}},
  \bibinfo{pages}{240} (\bibinfo{year}{1980}).

\bibitem[{\citenamefont{Mott and Massey}(1965)}]{Mott65}
\bibinfo{author}{\bibfnamefont{N.~F.} \bibnamefont{Mott}} \bibnamefont{and}
  \bibinfo{author}{\bibfnamefont{H.~S.~W.} \bibnamefont{Massey}},
  \emph{\bibinfo{title}{The Theory of Atomic Collisions}}
  (\bibinfo{publisher}{Oxford University Press, Oxford}, \bibinfo{year}{1965}).

\bibitem[{\citenamefont{Abramowitz and Stegun}(1965)}]{AS65}
\bibinfo{author}{\bibfnamefont{M.}~\bibnamefont{Abramowitz}} \bibnamefont{and}
  \bibinfo{author}{\bibfnamefont{I.~A.} \bibnamefont{Stegun}},
  \emph{\bibinfo{title}{Handbook of Mathematical Functions}}
  (\bibinfo{publisher}{Dover, New-York}, \bibinfo{year}{1965}).

\bibitem[{\citenamefont{Wait}(1967)}]{Wait}
\bibinfo{author}{\bibfnamefont{J.~R.} \bibnamefont{Wait}},
  \bibinfo{journal}{Can.\ J.\ Phys.} \textbf{\bibinfo{volume}{45}},
  \bibinfo{pages}{1861} (\bibinfo{year}{1967}).

\bibitem[{\citenamefont{Rayleigh}(1887)}]{Rayleigh1887}
\bibinfo{author}{\bibfnamefont{J.~W.~S.} \bibnamefont{Rayleigh}},
  \emph{\bibinfo{title}{The Theory of Sound \rm{reprinted by Dover in 1945}}}
  (\bibinfo{publisher}{Dover, New-York}, \bibinfo{year}{1887}).

\bibitem[{\citenamefont{Rayleigh}(1910)}]{Rayleigh1910}
\bibinfo{author}{\bibfnamefont{J.~W.~S.} \bibnamefont{Rayleigh}},
  \bibinfo{journal}{Phil.\ Mag.} \textbf{\bibinfo{volume}{20}},
  \bibinfo{pages}{1001} (\bibinfo{year}{1910}).

\bibitem[{\citenamefont{Berry et~al.}(1980)\citenamefont{Berry, Chambers,
  Large, Upstill, and Walmsley}}]{Berryetal80}
\bibinfo{author}{\bibfnamefont{M.}~\bibnamefont{Berry}},
  \bibinfo{author}{\bibfnamefont{R.}~\bibnamefont{Chambers}},
  \bibinfo{author}{\bibfnamefont{M.}~\bibnamefont{Large}},
  \bibinfo{author}{\bibfnamefont{C.}~\bibnamefont{Upstill}}, \bibnamefont{and}
  \bibinfo{author}{\bibfnamefont{J.}~\bibnamefont{Walmsley}},
  \bibinfo{journal}{Eur.\ J.\ Phys.} \textbf{\bibinfo{volume}{1}},
  \bibinfo{pages}{154} (\bibinfo{year}{1980}).

\bibitem[{\citenamefont{Roux et~al.}(1997)\citenamefont{Roux, de~Rosny, Tanter,
  and Fink}}]{Roux97}
\bibinfo{author}{\bibfnamefont{P.}~\bibnamefont{Roux}},
  \bibinfo{author}{\bibfnamefont{J.}~\bibnamefont{de~Rosny}},
  \bibinfo{author}{\bibfnamefont{M.}~\bibnamefont{Tanter}}, \bibnamefont{and}
  \bibinfo{author}{\bibfnamefont{M.}~\bibnamefont{Fink}},
  \bibinfo{journal}{Phys.\ Rev.\ Lett.} \textbf{\bibinfo{volume}{79}},
  \bibinfo{pages}{3170} (\bibinfo{year}{1997}).

\bibitem[{\citenamefont{Manneville et~al.}(2001)\citenamefont{Manneville, Roux,
  Tanter, Maurel, Fink, Bottausci, and Petitjeans}}]{Manneville01}
\bibinfo{author}{\bibfnamefont{S.}~\bibnamefont{Manneville}},
  \bibinfo{author}{\bibfnamefont{P.}~\bibnamefont{Roux}},
  \bibinfo{author}{\bibfnamefont{M.}~\bibnamefont{Tanter}},
  \bibinfo{author}{\bibfnamefont{A.}~\bibnamefont{Maurel}},
  \bibinfo{author}{\bibfnamefont{M.}~\bibnamefont{Fink}},
  \bibinfo{author}{\bibfnamefont{F.}~\bibnamefont{Bottausci}},
  \bibnamefont{and}
  \bibinfo{author}{\bibfnamefont{P.}~\bibnamefont{Petitjeans}},
  \bibinfo{journal}{Phys.\ Rev.\ E} \textbf{\bibinfo{volume}{63}},
  \bibinfo{pages}{036607} (\bibinfo{year}{2001}).

\bibitem[{\citenamefont{Berthet}(2001)}]{Berthet01}
\bibinfo{author}{\bibfnamefont{R.}~\bibnamefont{Berthet}},
  \emph{\bibinfo{title}{Interaction son-\'ecoulement, Ph.D. Thesis}}
  (\bibinfo{publisher}{Ecole Normale Sup\'erieure de Lyon},
  \bibinfo{year}{2001}).

\bibitem[{\citenamefont{Vivanco et~al.}(1999)\citenamefont{Vivanco, Melo,
  Coste, and Lund}}]{Vivanco99}
\bibinfo{author}{\bibfnamefont{F.}~\bibnamefont{Vivanco}},
  \bibinfo{author}{\bibfnamefont{F.}~\bibnamefont{Melo}},
  \bibinfo{author}{\bibfnamefont{C.}~\bibnamefont{Coste}}, \bibnamefont{and}
  \bibinfo{author}{\bibfnamefont{F.}~\bibnamefont{Lund}},
  \bibinfo{journal}{Phys.\ Rev.\ Lett.} \textbf{\bibinfo{volume}{83}},
  \bibinfo{pages}{1966} (\bibinfo{year}{1999}).

\bibitem[{\citenamefont{Coste et~al.}(1999)\citenamefont{Coste, Lund, and
  Umeki}}]{Coste99I}
\bibinfo{author}{\bibfnamefont{C.}~\bibnamefont{Coste}},
  \bibinfo{author}{\bibfnamefont{F.}~\bibnamefont{Lund}}, \bibnamefont{and}
  \bibinfo{author}{\bibfnamefont{M.}~\bibnamefont{Umeki}},
  \bibinfo{journal}{Phys.\ Rev.\ E} \textbf{\bibinfo{volume}{60}},
  \bibinfo{pages}{4908} (\bibinfo{year}{1999}).

\bibitem[{\citenamefont{Coste and Lund}(1999)}]{Coste99II}
\bibinfo{author}{\bibfnamefont{C.}~\bibnamefont{Coste}} \bibnamefont{and}
  \bibinfo{author}{\bibfnamefont{F.}~\bibnamefont{Lund}},
  \bibinfo{journal}{Phys.\ Rev.\ E} \textbf{\bibinfo{volume}{60}},
  \bibinfo{pages}{4917} (\bibinfo{year}{1999}).

\bibitem[{\citenamefont{Bernal et~al.}(2002)\citenamefont{Bernal, Coste, Lund,
  and Melo}}]{Bernal02}
\bibinfo{author}{\bibfnamefont{R.}~\bibnamefont{Bernal}},
  \bibinfo{author}{\bibfnamefont{C.}~\bibnamefont{Coste}},
  \bibinfo{author}{\bibfnamefont{F.}~\bibnamefont{Lund}}, \bibnamefont{and}
  \bibinfo{author}{\bibfnamefont{F.}~\bibnamefont{Melo}},
  \bibinfo{journal}{Phys.\ Rev.\ Lett.} \textbf{\bibinfo{volume}{89}},
  \bibinfo{pages}{034501} (\bibinfo{year}{2002}).

\bibitem[{\citenamefont{Neshev et~al.}(2001)\citenamefont{Neshev,
  Nepomnyashchy, and Kivshar}}]{Neshev01}
\bibinfo{author}{\bibfnamefont{D.}~\bibnamefont{Neshev}},
  \bibinfo{author}{\bibfnamefont{A.}~\bibnamefont{Nepomnyashchy}},
  \bibnamefont{and} \bibinfo{author}{\bibfnamefont{Y.~S.}
  \bibnamefont{Kivshar}}, \bibinfo{journal}{Phys.\ Rev.\ Lett.}
  \textbf{\bibinfo{volume}{87}}, \bibinfo{pages}{043901}
  (\bibinfo{year}{2001}).

\end{thebibliography}
\end{document}